\documentclass[%
 reprint,
superscriptaddress,
 amsmath,amssymb,
 aps,
]{revtex4-2}

\usepackage{amssymb,amstext,amsmath}
\usepackage{graphicx}
\usepackage{dcolumn}
\usepackage{bm} 
\usepackage{bbm} 
\usepackage{multirow}
\usepackage{bigstrut}
\usepackage{makecell,rotating}
\usepackage{mathrsfs}
\usepackage{booktabs}
\usepackage{threeparttable}
\usepackage{subfigure}
\usepackage{chngpage}
\usepackage{float}
\usepackage{color}
\usepackage{mathtools}
\usepackage{url}
\definecolor{darkred}{rgb}{0.75,0,0}
\definecolor{darkgreen}{rgb}{0,0.5,0}
\definecolor{darkblue}{rgb}{0,0,0.75}
\definecolor{darkorange}{rgb}{0.8,0.3,0}
\definecolor{dark}{rgb}{0,0,0}

\newcommand{\eq}[1]{Eq.~(\ref{eq:#1})}
\newcommand{\fig}[1]{Fig.~\ref{fig:#1}}

\begin{document}

\preprint{APS/123-QED}

\title{Promoting collective cooperation through temporal interactions}

\author{Yao Meng}
\affiliation{%
	Center for Systems and Control, College of Engineering, Peking University, Beijing 100871, China}

\author{Alex McAvoy}
\affiliation{School of Data Science and Society, University of North Carolina at Chapel Hill, Chapel Hill, NC 27599, USA}
\affiliation{Department of Mathematics, University of North Carolina at Chapel Hill, Chapel Hill, NC 27599, USA}

\author{Aming Li}
\thanks{amingli@pku.edu.cn}%
\affiliation{%
	Center for Systems and Control, College of Engineering, Peking University, Beijing 100871, China}
\affiliation{
	Center for Multi-Agent Research, Institute for Artificial Intelligence, Peking University, Beijing 100871, China}

\date{\today}

\begin{abstract}
Collective cooperation drives the dynamics of many natural, social, and economic phenomena, making understanding the evolution of cooperation with evolutionary game theory a central question of modern science.
Although human interactions are best described as complex networks, current explorations are limited to static networks where interactions represented by network links are permanent and do not change over time. 
In reality, human activities often involve temporal interactions, where links are impermanent, 
and understanding the evolution of cooperation on such ubiquitous temporal networks is an open question.
Here, we present a general framework for systematically analyzing how collective cooperation evolves on any temporal network, which unifies the study of evolutionary game dynamics with dynamic and static interactions. 
We show that the emergence of cooperation is facilitated by a simple rule of thumb: hubs (individuals with many social ties) should be temporally deprioritized in interactions. 
We further provide a quantitative metric capturing the priority of hubs, which we utilize to orchestrate the ordering of interactions to best promote cooperation on empirical temporal networks. 
Our findings unveil the fundamental advantages conferred by temporal interactions for promoting collective cooperation, which transcends the specific insights gleaned from studying traditional static snapshots.
\end{abstract}

\maketitle
\section{Introduction}
Explaining the prevalence of altruistic behaviors among self-interested individuals has been a central topic in theoretical biology, dating back to the seminal work of Hamilton \cite{hamilton:JTB:1964a,hamilton:JTB:1964b}. In recent decades, researchers have sought to explain the emergence of cooperation in finely-structured populations using evolutionary game theory \cite{Nowak92Nature,hofbauer1998evolutionary,nowak2004emergence,Sigmund2010,nowak2010evolutionary}, where the extensive variety of interpersonal interactions that humans engage in on a daily basis are best described with networks \cite{hauert2004spatial,lieberman2005evolutionary,santos2005scale,ohtsuki2006simple,Ohtsuki2007symmetry,taylor2007evolution,Santos2008Social,szolnoki2009topology,perc2013evolutionary,maciejewski2014evolutionary,allen2017evolutionary,hilbe2018evolution,Su2019,levin2020collective,li2020evolution,chen2022searching}. 
A major strand of ongoing research concerns understanding how cooperation self-organizes through a combination of network structure and the mechanism of behavioral transmission, as well as how empirical data on these two factors can be incorporated into collective population dynamics.

Although theoretical research in evolutionary dynamics has (partially) elucidated the role of network structure in the spread of social traits, many such studies rely on the key assumption that the underlying network is static. 
Within this paradigm, heterogeneous structures---in which individuals can have different numbers of neighbors---have long been recognized as important for capturing realistic populations in modeling approaches. 
Qualitatively novel dynamics and properties emerge in empirical systems modeled by heterogeneous networks, which are absent from homogeneous topologies wherein all individuals have the same number of neighbors.
For example, in public goods games, heterogeneous networks can lead to heightened wealth inequality, following a power law distribution on scale-free networks \cite{Santos2008Social}. Such networks can also promote the spread of inefficient prosocial behaviors, in which the costs vastly exceed the benefits \cite{mcavoy:NHB:2020,ventura:PS:2022}.

In reality, (static) heterogeneous networks may fall short of capturing the complexity of human social interactions. Numerous interpersonal exchanges are dynamic, characterized by networks that vary over time \cite{gelardi:PRSB:2021}.
Instances include electronic communication through both email and question-and-answer websites (e.g., Stack Overflow) \cite{paranjape:ICWSDM:2017}, as well as face-to-face interactions like those in schools \cite{student2012} and workplaces \cite{office2013}. 
Although precise timestamped data on temporal networks is not always readily available, randomness in interaction patterns across species (e.g., due to weather, seasonality, travel, or migration) suggests that temporally varying contact networks are integral components of social evolution and should be considered in modeling efforts.

Static networks, which are widely studied, involve the aggregation of dynamic interactions over time, which inevitably results in the loss of information about when interactions occur \cite{Renaud2016}.
Moreover, it is crucial to emphasize that the temporal aspect of interactions plays a substantial role in various dynamic processes \cite{masuda2013temporal,scholtes2014causality,li2017fundamental,karsai2021}.
However, the exploration of cooperation on temporal networks has, until now, been confined to numerical simulations and approximations \cite{akcay:NC:2018,li2020evolution,cardillo2014evolutionary}, lacking a general framework for unifying the study of  evolutionary dynamics with temporal and static interactions.
Here, we derive a mathematical condition for when cooperation emerges on temporal networks, which includes an explicit description of how network changes are incorporated. 
Our findings reveal that prioritizing individuals with fewer social ties is crucial for the efficient spread of cooperative behaviors over time.
As applications, we design optimal temporal orderings of interactions on both synthetic and empirical datasets. Our theoretical results are general and are developed with future applications in mind, as more empirical data about temporal interactions becomes available.

\section{Model}
\begin{figure*}
	\centering
	\includegraphics[width=0.95 \textwidth]{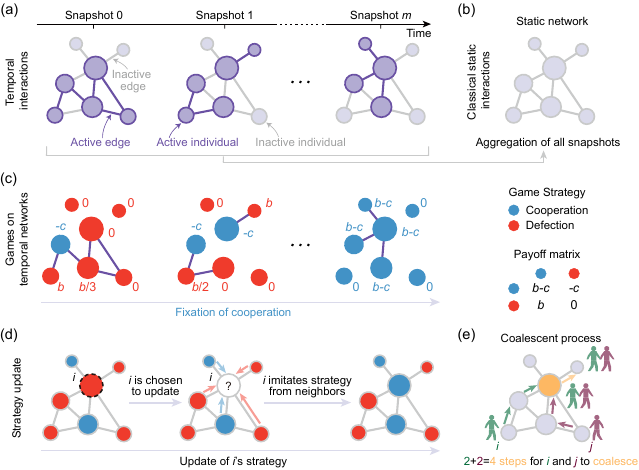}
	\caption{Evolutionary games with temporal interactions.
		(a) Temporal interactions between individuals over time are represented by a snapshot sequence, where the active edges at each time step are marked in dark purple, and an individual is active if it has at least one active edge. 
		(b) The static network aggregated from all snapshots in the temporal network captures the social relationships among individuals. 
		Edges or nodes that appear in the static network but are inactive in the snapshots of temporal networks are marked in gray.
		(c) Individuals choose either cooperation (blue) or defection (red) as their strategy in the game. At each time step, each individual, $i$, plays the donation game with all current neighbors and obtains an average payoff, $f_{i}$, where a cooperator pays a cost $c$ to provide a benefit $b$ to its opponent, and defectors provide no benefit and pay no cost. 
		(d) After each time step of interactions, a randomly selected individual, $i$ (dashed circle), updates its strategy following an imitation mechanism. The neighbors then compete to be imitated, and the probability that $i$ copies the behavior of neighbor $j$ is proportional to $j$'s fitness, $F_{j}=1+\delta f_{j}$.
		Starting from a single cooperator, the evolutionary process ends when cooperators either fix or go extinct.
		(e) Random walkers starting from $i$ and $j$ meet on the yellow node and then continue to make a single random walk, in this example coalescing in four time steps, with each taking two steps before they meet. We take $\mathbb{P}_{\left(i,j\right)}\left[\tau =m\right]$ to represent the probability that the lineages leading to $i$ and $j$ coalesce at $m$ time steps.
	} \label{fig:illustration}
\end{figure*}

We consider games with temporal interaction patterns in a population of $N$ individuals, where the network structures can vary at each time step (\fig{illustration}a). 
The aggregated social relationships are represented by a static network, obtained by combining all interaction snapshots over time. This structure is also known as the ``replacement'' network \cite{Ohtsuki2007symmetry}, which captures who can imitate whom during the evolutionary process (\fig{illustration}b).

Individuals choose either cooperation or defection. At each time step, each individual, $i$, plays the game pairwise with its interaction partners and obtains an average payoff, $f_{i}$. A cooperator pays a cost, $c$, to provide a benefit, $b$, to its opponent. Defectors pay nothing and provide no benefit (\fig{illustration}c). An individual, $i$, is then chosen uniformly at random from the population to imitate a strategy from one of its neighbors, $j$, with probability proportional to the fitness of $j$. This ``fitness'' is defined as $F_{j}=1+\delta f_{j}$, where $\delta >0$ captures the intensity of selection (\fig{illustration}d).

Since individuals are chosen uniformly at random for strategy evaluation, a novel behavior (e.g., cooperate in an all-defector population or defect in an all-cooperator population) arises in location $i$ with probability $1/N$. Once the mutant appears, the population updates until the mutant type either fixes or goes extinct, and then another mutant is introduced. Thus, to quantify the evolutionary success of cooperators, we consider the probability that a mutant type takes over when placed uniformly at random within a population of residents. We denote by $\rho_{C}$ and $\rho_{D}$ the fixation probabilities of cooperators and defectors, respectively. Under neutral drift, temporal interactions have no effect on these quantities since payoffs are then immaterial and both $\rho_{C}$ and $\rho_{D}$ are equal to $1/N$. Under weak-selection, meaning $0<\delta\ll 1$, cooperation is said to be favored relative to defection if $\rho_{C}>\rho_{D}$, which will be the main focus of this study.

\section{Results}

\subsection{Theoretical condition for the evolution of cooperation on temporal networks}
The evolution of cooperation on temporal interactions depends on how strategies spread among individuals, which can be studied using an ancestral process on the replacement network (\fig{illustration}e). The idea is that, since we are concerned with weak selection, neutral dynamics should yield useful insights into the assortment of behaviors, which can then be used to understand the effects of small selection strengths. We establish this intuition formally in Supplementary Information. Under neutral drift, one can look backward in time to understand correlations between behaviors of individuals in the present generation, based on the fact that correlations arise from sharing common ancestors.

Before stating our main theoretical finding, we need to introduce some technical notation. For $i\neq j$, let $\mathbb{P}_{\left(i,j\right)}\left[T^{\textrm{coal}}\leqslant t\right]$ denote the probability that the lineages leading to $i$ and $j$ coalesce at most $t\geqslant 0$ time steps into the past \cite{cox1989coalescing,kingman1982coalescent}. 
At time $t$, individual $i$ has $I_{i}\left(t\right)\coloneqq\sum_{j=1}^{N}I_{ij}\left(t\right)$ interaction partners in the current interaction snapshot, where $I_{ij}\left(t\right) =I_{ji}\left(t\right) =1$ indicates that there is an interaction between players $i$ and $j$ at time $t$ (and $I_{ij}\left(t\right) =I_{ji}\left(t\right) =0$ otherwise). 
The probability of moving from $i$ to $j$ in one step of a random walk on the snapshot at time $t$ is $q_{ij}\left(t\right)\coloneqq I_{ij}\left(t\right) /I_{i}\left(t\right)$ if $I_{i}\left(t\right) >0$ (and $q_{ij}\left(t\right)\coloneqq 0$ otherwise).
Similarly, the probability of moving from $i$ to $j$ in one step of a random walk on the replacement network is $p_{ij}\coloneqq w_{ij}/w_{i}$, where $w_{i}=\sum_{j=1}^{N}w_{ij}$ and $w_{ij}=w_{ji}=1$ if players $i$ and $j$ interact at least once over time ($w_{ij}=w_{ji}=0$ otherwise). Note that coalescent times depend on the replacement network only, and $\mathbb{P}_{\left(i,j\right)}\left[T^{\textrm{coal}}\leqslant t\right]$ can be calculated using the single-step probabilities, $p_{ij}$ (see Supplementary Information). Finally, let $\pi_{i}\coloneqq w_{i}/\sum_{j=1}^{N}w_{i}$ be the reproductive value \cite{allen2017evolutionary} of $i$, which defines a distribution over locations in the replacement network, favoring more highly-connected nodes.

At time $t$, consider the derived benefit $\mathcal{B}_{n}\left(t\right)\coloneqq b\sum_{i,j,k=1}^{N}\pi_{i}p_{ij}^{\left(n\right)} q_{jk}\left(t\right)\mathbb{P}_{\left(i,k\right)}\left[T^{\textrm{coal}}\leqslant t\right]$, where $p_{ij}^{\left(n\right)}$ generalizes $p_{ij}$, representing the $n$-step probability of moving from $i$ to $j$. 
The idea is to place a cooperator at location $i$ with probability $\pi_{i}$. We then take an $n$-step random walk to location $j$ on the replacement network. Once at $j$, we take one step of a random walk on the interaction network at time $t$, to location $k$. Since $i$ is a cooperator, the probability that $k$ descends from a common cooperator at time $t$ is $\mathbb{P}_{\left(i,k\right)}\left[T^{\textrm{coal}}\leqslant t\right]$. Each such $k$ contributes a benefit to $j$, so $\mathcal{B}_{n}\left(t\right)$ represents the expected benefit to the individual at the end of an $n$-step random walk from a cooperator. Analogously, the time-$t$ derived cost to an individual at the end of an $n$-step random walk from a cooperator is $\mathcal{C}_{n}\left(t\right)\coloneqq c \sum_{i,j=1}^{N}\pi_{i}p_{ij}^{\left(n\right)} q_{j}\left(t\right)\mathbb{P}_{\left(i,j\right)}\left[T^{\textrm{coal}}\leqslant t\right]$, where $q_{j}\left(t\right):=\sum_{k=1}^{N} q_{jk}(t)$ indicates whether individual $j$ has interactions at time step $t$. 
The primary difference between $\mathcal{B}_{n}\left(t\right)$ and $\mathcal{C}_{n}\left(t\right)$, apart from the factors $b$ and $c$, is that $\mathcal{C}_{n}\left(t\right)$ depends on $\mathbb{P}_{\left(i,j\right)}\left[T^{\textrm{coal}}\leqslant t\right]$ instead of $\mathbb{P}_{\left(i,k\right)}\left[T^{\textrm{coal}}\leqslant t\right]$, owing to the fact that, whenever $j$ is a cooperator, $j$ incurs a cost for each interaction partner $k$ at time $t$.

Our main result says, roughly, that cooperators are favored over defectors whenever
\begin{equation}
-\sum_{t=0}^{\infty} \mathcal{C}_{0}\left(t\right) +\sum_{t=0}^{\infty} \mathcal{B}_{0}\left(t\right) > -\sum_{t=0}^{\infty} \mathcal{C}_{2}\left(t\right) + \sum_{t=0}^{\infty} \mathcal{B}_{2}\left(t\right) .\label{eq:intuition}
\end{equation}
Intuitively, this condition means that cooperators are favored if they have, on average, a higher payoff than a random individual two steps away in the replacement network. At a high level, this is the same intuition for the corresponding condition on static networks \cite{allen2017evolutionary}, with the reasoning being that cooperators compete with two-step neighbors to have their strategy imitated by a common neighbor. However, the condition of Allen et al. \cite{allen2017evolutionary} cannot be evaluated on temporal networks. The reason the intuition in \eq{intuition} is ``rough'' is that summing $\mathcal{B}_{n}\left(t\right)$ and $\mathcal{C}_{n}\left(t\right)$ over all $t$ generally leads to divergent series. To mitigate this issue, we replace $\mathbb{P}_{\left(i,j\right)}\left[T^{\textrm{coal}}\leqslant t\right]$ by $1-\mathbb{P}_{\left(i,j\right)}\left[T^{\textrm{coal}}>t\right]$ and cancel out the common factor in \eq{intuition}, which leads to the finding that selection favors cooperators whenever
\begin{equation}
	\frac{b}{c} > \frac{\sum_{t=0}^{\infty}\sum_{i,j=1}^{N}\pi_{i} p_{ij}^{\left(2\right)} q_{j}\left(t\right) \mathbb{P}_{\left(i,j\right)}\left[T^{\textrm{coal}} > t \right]}{\left(\substack{\sum_{t=0}^{\infty}\sum_{i,j,k=1}^{N}\pi_{i} p_{ij}^{\left(2\right)} q_{jk}\left(t\right)\mathbb{P}_{\left(i,k\right)}\left[T^{\textrm{coal}} > t \right] \\-\sum_{t=0}^{\infty}\sum_{i,j=1}^{N}\pi_{i} q_{ij}\left(t\right)\mathbb{P}_{\left(i,j\right)}\left[T^{\textrm{coal}} > t \right]}\right)} .
\label{eq:condition}
\end{equation}
In Supplementary Information, we give a formal proof of this result and demonstrate how to evaluate it. 

\begin{figure*}
	\centering
	\includegraphics[width=0.95 \textwidth]{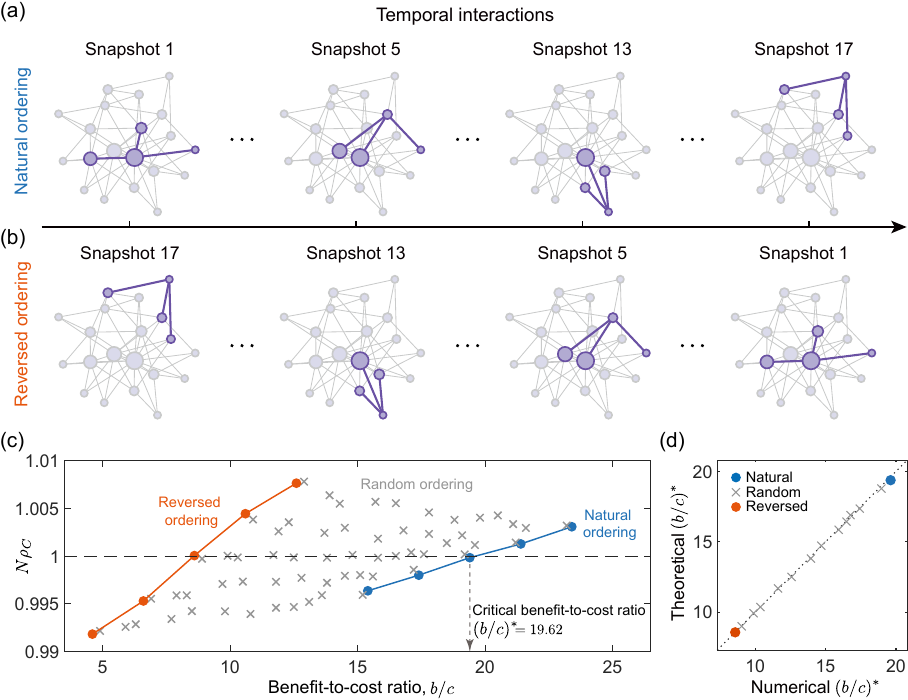}
	\caption{Effect of interaction ordering on the emergence of cooperation.
		(a) We generate temporal interactions under a preferential attachment (Barab\'asi-Albert) model \cite{Barabasi1999a}, where a node is added to the network and connected to $m=3$ different existing nodes at each step. We take the newly added edges and nodes of the both ends as the interaction network (purple) in each snapshot, which lasts for $N=20$ time steps. We illustrate the interactions in snapshots $1$, $5$, $13$, and $17$. The replacement network (gray) has $N$ nodes and is exactly the generated network. Interactions with reversed time-ordering are shown in (b). 
		(c) We show the fixation probability of cooperation ($\rho_{C}$) as a function of the benefit-to-cost ratio ($b/c$) for natural (blue circle), random (gray cross), and reversed (orange circle) ordering. The critical benefit-to-cost ratio, $\left(b/c\right)^{\ast}$, for a given network occurs when the corresponding curve intersects the horizontal line ($\rho_{C}=1/N$), representing neutral drift. $\left(b/c\right)^{\ast}$ for the natural ordering is highlighted in (c), which is slightly less than $20$. (d) The simulation results are in good agreement with theoretical predictions calculated using \eq{condition}.
	} \label{fig:preferential_attachment}
\end{figure*}

\subsection{The effect of chronological ordering on evolutionary dynamics}
We first study temporal interactions generated randomly using the Barab\'asi-Albert model \cite{Barabasi1999a}, in which the network is constructed by linking each new vertex to $m$ existing vertices, with probability proportional to the degrees of the existing vertices. The nodes that are added first tend to have large degree, while those added later have fewer neighbors. Due to the way in which Barab\'asi-Albert networks are constructed, there is a natural choice for the ordering of interactions. \fig{preferential_attachment}a shows the snapshots of interactions over time, where the active node plays games pairwise with the $m$ neighbors it gets connected to, and the replacement of strategies occurs on the underlying aggregated network. Each snapshot lasts for $N$ time steps (equal to the population size), with each individual updating its strategy once, on average, during the snapshot.

We find that the natural ordering tends to result in the largest critical benefit-to-cost ratio, $\left(b/c\right)^{\ast}$, meaning this ordering is least favorable to the evolution of cooperation. We can reverse the ordering of the temporal interactions in \fig{preferential_attachment}b, where the individuals with few connections (small degree) on the replacement network are prioritized for early interactions, while individuals with large degree interact later. In this reversed scheme, we find a marked drop in $\left(b/c\right)^{\ast}$ relative to the natural ordering (\fig{preferential_attachment}c). These two orderings tend to be extremes in this model, with random orderings yielding critical ratios in between (\fig{preferential_attachment}c,d). The numerical results are in good agreement with the theoretical prediction of \eq{condition} (\fig{preferential_attachment}d).

\subsection{Promoting cooperation with temporal interactions}
To better explore how the temporal ordering of interactions affects the evolution of cooperation, we can simplify \eq{condition} using a mean-field approximation. It is known that coalescence times are related to meeting times of independent random walks on the replacement network \cite{allen2017evolutionary}. Let $\tau$ denote the meeting time for such random walks, with $\mathbb{P}_{\left(i,j\right)}\left[\tau =m\right]$ being the probability that two random walks starting at $i$ and $j$, respectively, meet after exactly $m$ time steps, where in each time step, exactly one walk is selected (uniformly at random) to take a step (\fig{illustration}e). 
We show the calculation of this probability, as well as its relationship to coalescence times, in Supplementary Information. While $\mathbb{P}_{\left(i,j\right)}\left[\tau =m\right]$ generally depends on $i$ and $j$, here we approximate these quantities using an average over all starting locations, which we denote by $\mathbb{P}\left[\tau =m\right]$.

We introduce two important variables to capture the interaction structure at time $t$. Let $q\left(t\right)\coloneqq\sum_{i=1}^{N}\pi_{i}q_{i}\left(t\right)$. Since $q_{i}\left(t\right) =1$ when $I_{i}\left(t\right) >0$ and $0$ otherwise, $q\left(t\right)$ is equal to the sum of reproductive value over all nodes that participate in at least one interaction at time $t$. Let $\ell\left(t\right)\coloneqq\sum_{i,j=1}^{N}\pi_{i}\left(\frac{1}{2}p_{ij}q_{ji}\left(t\right) +\frac{1}{2}q_{ij}\left(t\right) p_{ji}\right)$ be a modified two-step return probability, which captures the effects of overlap in the replacement and interaction networks at time $t$. Intuitively, we start at location $i$ with probability $\pi_{i}$, and then ask about the mean probability of moving first from $i$ to a neighbor and then back to $i$ in the subsequent step. With probability $1/2$, the first step is taken in the replacement network and the second in the interaction network. Otherwise, the interaction network comes first and the replacement network follows.

We now define accumulated time-averaged reproductive value, $Q$, and overlap, $L$, as $Q\coloneqq\sum_{m=1}^{\infty}\sum_{T=0}^{m-1}q\left(\left\lfloor TN/2\right\rfloor\right)\mathbb{P}\left[\tau =m\right]$ and $L\coloneqq\sum_{m=1}^{\infty}\sum_{T=0}^{m}\ell\left(\left\lfloor TN/2\right\rfloor\right)\mathbb{P}\left[\tau =m\right]$, respectively. Using these averages, we can approximate \eq{condition} using the ratio
\begin{equation}
\left(\frac{b}{c}\right)^{\ast} \approx \frac{Q+\varepsilon_{c}}{L-2\overline{q}+\varepsilon_{b}} ,\label{eq:approx_bcr_temporal}
\end{equation}
where $\overline{q}:=\sum_{m=0}^{\infty}q\left(\left\lfloor mN/2\right\rfloor\right)\mathbb{P}\left[\tau =m\right]$ represents the time-averaged reproductive value, which is positively related to $Q$ (Supplementary Fig.~S1b).
Here, $\varepsilon_{c}$ and $\varepsilon_{b}$ are negligible compared to the other quantities (see Supplementary Information). This approximation holds with remarkable accuracy (Supplementary Figs.~S1a and S2). 
Since $\mathbb{P}\left[\tau =m\right]$ decays exponentially in $m$ (Supplementary Fig.~S1c), \eq{approx_bcr_temporal} implies that early interaction structures are weighted more in the critical ratio than later snapshots. In particular, lower values of $\left(b/c\right)^{\ast}$ are achieved by structures for which $L$ is large relative to $Q$ (Supplementary Fig.~S1d), so interactions with small $q\left(t\right)$ and large $\ell\left(t\right)$ should occur early.

By its definition, $q\left(t\right)$ is small whenever the individuals involved in interactions at time $t$ have relatively low degree in the replacement network (or, if all degrees are comparable, only a small fraction of individuals participate in interactions). 
Therefore, roughly speaking, the critical ratio in \eq{approx_bcr_temporal} is lower when interactions between individuals of low degree in the replacement network occur earlier in time.
$\ell\left(t\right)$, on the other hand, is generally larger when the two networks have greater overlap. 
However, it is not exactly obvious that there is an optimizer for $\ell\left(t\right)$ when the replacement network is held fixed. In the case that one degree is much larger than the others, there is a tension between larger $\ell\left(t\right)$ and smaller $q\left(t\right)$, since now the node of largest degree should participate in the interaction to maximize $\ell\left(t\right)$, which, in turn, increases $q\left(t\right)$.

\subsection{Implications for static networks}
We now consider what our analysis says about static interaction networks that do not vary in time. Our examples so far have involved replacement networks obtained from the interaction structure; as such, the assumption that the interaction network is static then implies that the two networks are identical. However, the main theoretical expressions in Supplementary Information allow the replacement and interaction networks to be independent. This scenario, of static but distinct interaction and replacement networks, has been considered in several previous studies \cite{taylor2007evolution,Ohtsuki2007symmetry,debarre:NC:2014}. Here, we consider the relationship between static and temporal networks conducive to the evolution of cooperation.

For static interactions where the structures do not vary through time, we denote $q\left(t\right)$ and $\ell\left(t\right)$ in \eq{approx_bcr_temporal} by $q$ and $\ell$, respectively, for simplicity. The critical ratio for static interactions is then
\begin{equation}
\left(\frac{b}{c}\right)^{\ast} \approx \frac{\overline{\tau}q + \varepsilon_{c}}{\left(1+\overline{\tau}\right)\ell -2q +\varepsilon_{b}} ,
\label{eq:approx_bcr_static}
\end{equation}
where $\overline{\tau}=\sum_{m=1}^{\infty}m\mathbb{P}\left[\tau =m\right]$ represents the average expected coalescence time over any pair of nodes on the replacement network. Therefore, the interaction structure with a small $q$ and a large $\ell$ leads to a lower critical ratio for static interactions. This finding says that, in a temporal network, if the interaction snapshots appearing early in time are such that they would promote cooperation when held static, then cooperation has a lower barrier to emergence in the temporal setting.

\begin{figure*}
	\centering
	\includegraphics[width=0.95\textwidth]{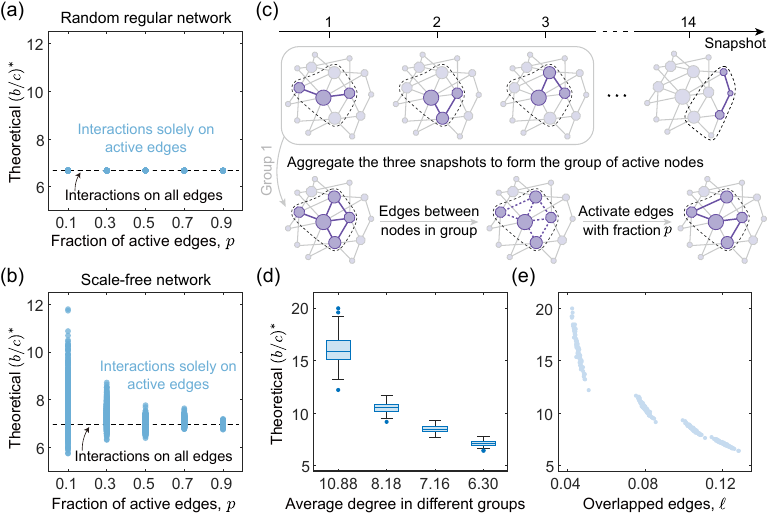}
	\caption{Favorable interaction structures on static networks.
		We present $\left(b/c\right)^{\ast}$ on interaction structures generated by randomly choosing edges at fraction $p$ from random regular and scale-free networks in (a) and (b) respectively, which form the replacement networks for the evolutionary process. Here, each dot corresponds to an interaction structure, and $100$ samples are plotted over each value of $p$. 
		(c) We divide the subsequent snapshots in a temporal network generated under preferential attachment into four groups, where each node group (dashed bound) is formed by the active nodes (purple) on the corresponding aggregated interaction network.
		Interaction edges (dashed line) between nodes (purple) in each group are selected with probability $p=0.5$ from the underlying replacement networks. 
		(d) For the four groups designed from a temporal network with $100$ nodes using the methods in (c), the groups with more hubs yield higher $\left(b/c\right)^{\ast}$ than those with more leaves. 
		We calculate $\left(b/c\right)^{\ast}$ as a function of $\ell$ in (e) for interactions between nodes in each group with $100$ samples, where the fewer interaction partners a hub has, the lower the value of $\left(b/c\right)^{\ast}$ is. 
	}\label{fig:static_temporal}
\end{figure*}

We now explore how each interaction snapshot plays a role in the evolution of cooperation in the scenario of static interactions. We take random regular and scale-free networks as replacement networks, and generate the interaction networks by randomly choosing edges at fraction $p$. The active edges, together with the nodes connected by the active edges, form the interaction networks. We find that the critical threshold of an interaction structure constructed from random regular networks is nearly unaffected by the fraction of active edges (\fig{static_temporal}a). In contrast, the interaction structure strongly influences $\left(b/c\right)^{\ast}$ on scale-free networks, especially at small $p$ (\fig{static_temporal}b).

\begin{figure*}
	\centering
	\includegraphics[width=0.95\textwidth]{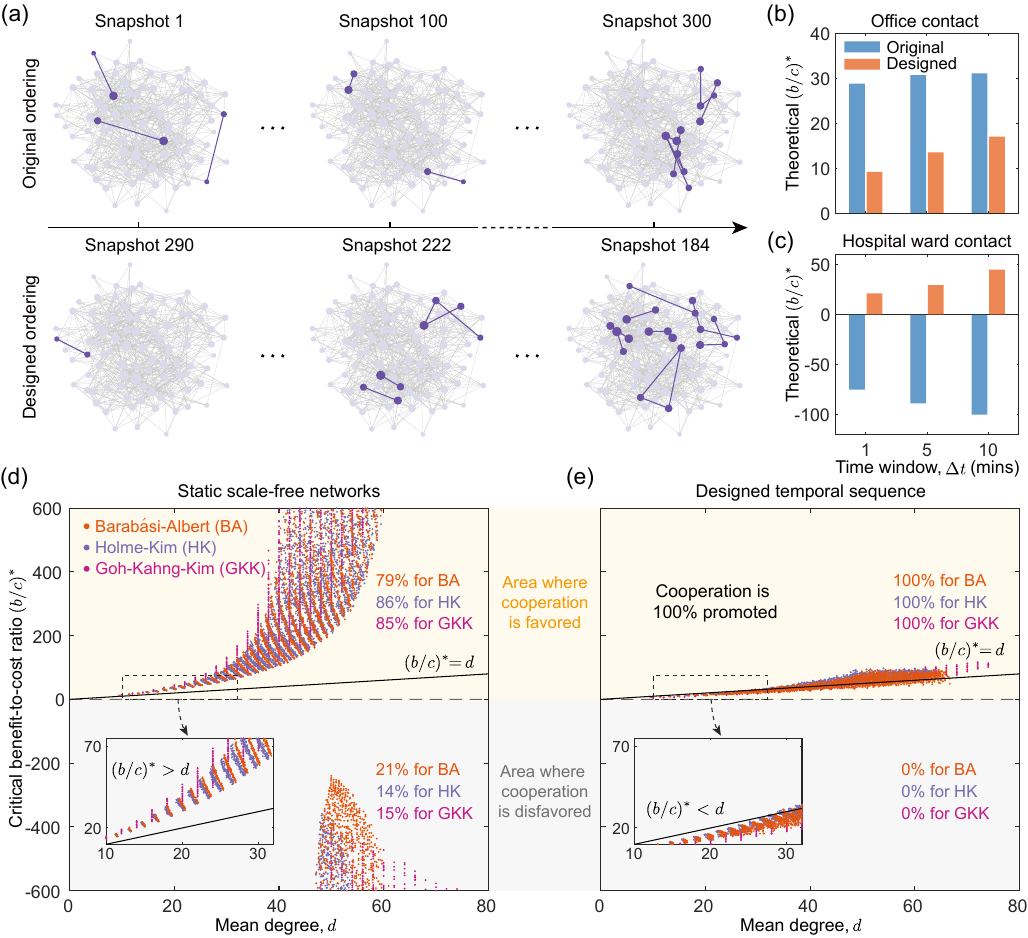}
	\caption{Designing temporal interactions on empirical datasets and replacement networks. 
		(a) For empirical temporal networks describing face-to-face contacts in an office building \cite{office2013}, we present the natural temporal interaction patterns (top panel) and our designed order of snapshots (bottom panel). The corresponding temporal networks are constructed by aggregating interactions with time window of $\Delta t=10$ minutes. 
		We calculate the theoretical value of critical threshold $\left(b/c\right)^{\ast}$ on temporal interactions with both natural and designed ordering for two empirical datasets collected from contacts in (b) an office \cite{office2013} and (c) a hospital \cite{hospital2013}. 
		For different values of the time window, $\Delta t$, our designed ordering of interactions facilitates the emergence of cooperation. 
		(d) Scatter plot of the critical threshold, $\left(b/c\right)^{\ast}$, based on mean degree, $d$, on Barab\'asi-Albert (BA) \cite{Barabasi1999a}, Holme-Kim (HK) \cite{holme2002growing}, and Goh-Kahng-Kim (GKK) \cite{goh2001universal} scale-free networks.
		Each type of network has negative $\left(b/c\right)^{\ast}$, which implies that spite can be favored and there is no possibility for those networks to favor cooperation.
		(e) With our designed temporal interactions, all scale-free networks have positive $\left(b/c\right)^{\ast}$, which is approximately equal to the average degree ($d$) of the network.
		All structures have $10^4$ realizations, with the number of nodes $100 \leqslant N \leqslant 150$ and mean degree $10 \leqslant d \leqslant 100$ for each network. 
	} \label{fig:designing}
\end{figure*}

To explore the effects of which nodes are involved in interactions, we divide the snapshots in a temporal network generated under preferential attachment into four groups with generally equal number of snapshots (\fig{static_temporal}c).
This naturally leads to qualitative differences between the groups: the group with earlier snapshots tends to have more hubs, which results in a higher value of $q$ in \eq{approx_bcr_static}, i.e., a larger average degree of nodes within the group.
To further determine the effect caused by the overlap between the interaction and replacement network, given the same group of active individuals, we form the interaction network from the replacement network (the aggregated scale-free network) by randomly choosing edges between the nodes within each group, with probability $p=0.5$.
Interactions between nodes of small degree lead to a lower value of $\left(b/c\right)^{\ast}$, while interactions between the group with more hubs generally impede the emergence of cooperation (\fig{static_temporal}d). $\left(b/c\right)^{\ast}$ decreases when the degree of nodes who have interactions decreases. For given nodes in the static interaction networks, $\left(b/c\right)^{\ast}$ decreases when $\ell$ increases within each group of nodes (\fig{static_temporal}e). Moreover, for a given interaction between $i$ and $j$, $\ell$ will be larger with $\left(w_{i}-w_{j}\right)\left(I_{i}-I_{j}\right) <0$ than vice versa, where $I_{i}$ is the number of interaction partners of node $i$. In other words, the individual with fewer social connections should be more involved in interactions.

\subsection{Designing optimal temporal interactions}
Using the intuition gathered so far, we now ask whether we can better promote cooperation by reordering the sequence of snapshots. Here, we construct temporal interactions on two empirical datasets \cite{office2013,hospital2013} by aggregating social contacts over time windows of length $\Delta t$. We let $\mathcal{H}\coloneqq q/\ell$ be the priority of the hubs, which represents the degree to which hubs are involved in interactions. Interaction structures with larger $\mathcal{H}$ should occur later. \fig{designing}a illustrates the original interaction sequence, and the sequence arising when interactions are arranged in ascending ordering of $\mathcal{H}$. In cases for which $\left(b/c\right)^{\ast}>0$ (meaning cooperation can evolve at all), we find that the reordered sequence results in a lower value of $\left(b/c\right)^{\ast}$. This finding is illustrated in \fig{designing}b for different values of $\Delta t$. Remarkably, when $\left(b/c\right)^{\ast}<0$ in the original network (which means that spite is favored), the reordered sequence results in a positive value of $\left(b/c\right)^{\ast}$, presenting a new avenue for cooperation to evolve (\fig{designing}c). In both scenarios, the designed snapshot ordering enhances the emergence of cooperation, relative to the original interaction sequence.

In many practical situations, the exact timing of the interactions is unknown. By taking each edge on the static network as an interaction snapshot, the temporal network can be arranged in ascending ordering of hub priority, $\mathcal{H}$. \fig{designing}d,e present the critical threshold $\left(b/c\right)^{\ast}$ on designed temporal interactions based on three different types of scale-free networks. We find that although sparse scale-free networks require a high critical ratio, with $\left(b/c\right)^{\ast}$ larger than the mean degree $d$ (\fig{designing}d), orchestrated temporal interactions on scale-free networks are more favorable for cooperation, with $\left(b/c\right)^{\ast}<d$ (\fig{designing}e). Furthermore, $100\%$ of scale-free networks with orchestrated temporal interactions have positive values of $\left(b/c\right)^{\ast}$ and present significant advantages for altruistic behaviors (\fig{designing}e). Our results demonstrate the striking role of temporal interactions on highly heterogeneous networks, especially in scenarios where the corresponding static (aggregated) networks are dense.

\section{Discussion}
A growing body of research has shown that network structures, both static and time-varying, can promote the emergence of prosocial behaviors \cite{santos2005scale,ohtsuki2006simple,allen2017evolutionary}. Our findings provide a unifying analytical understanding of the evolution of cooperation with temporal interactions and how it relates to evolutionary dynamics on individual snapshots. We explicitly uncover that the presence of hubs impedes the evolution of cooperation in static networks, and that these hubs should be temporally deprioritized in time-varying interactions to promote cooperation.

Previous studies on asymmetric regular interaction and replacement structures suggest that the optimal threshold is reached when the interaction and replacement networks are identical \cite{Ohtsuki2007symmetry}. 
Surprisingly, our results reveal that for heterogeneous interaction and replacement, maximum overlap between these two networks is not optimal. Cooperation is difficult to maintain on high-degree nodes in the replacement network, as neighboring defectors threaten to change their behavior. In contrast, the more that small-degree nodes are involved in interactions, the more likely they are to copy strategies from the nodes they interact with, which eventually drives the emergence of cooperative clusters among non-hubs. When expanding the results to temporal interactions, we find that the threshold for favoring cooperation on heterogeneous networks is reduced by allowing individuals with fewer social ties to become more involved in interactions earlier on.

One promising application of our finding is to the design of interaction sequences in practical scenarios. The empirical temporal networks \cite{office2013,hospital2013} studied here illustrate interpersonal interactions at each time. Our work provides a concise and efficient metric to arrange interaction sequences---sorting the sequence in the ascending order of hub priority, $\mathcal{H}$, for each snapshot. For instance, meeting times between different teams in an office \cite{office2013} can be rearranged to maximize cooperative behaviors among team members. Hospitals \cite{hospital2013} can arrange and prioritize some designated patients for consultation in their scheduling. By arranging the sequence of interactions according to the guidelines provided here, cooperation can be significantly promoted (\fig{designing}b, c, Supplementary Fig.~S3).

In empirical social systems, heterogeneous networks are widespread. Our findings break through the limitations of studies on static networks in explaining the evolution of cooperation on heterogeneous networks. Previous results have shown that heterogeneous networks require a higher threshold for favoring cooperation than homogeneous networks \cite{ohtsuki2006simple,allen2017evolutionary,fotouhi2019evolution}. By offering design guidelines for temporal interactions, we find heterogeneous networks with our orchestrated temporal interactions can greatly facilitate cooperation compared to homogeneous networks, regardless of whether the network is sparse or dense (Supplementary Fig.~S4). Our theory thus extends insights into the evolution of cooperation to temporal networks, which is and will continue to be a focal topic in modern science.

%

\end{document}